\documentclass[aps,prl,showpacs,superscriptaddress,twocolumn]{revtex4}% ,longbibliography
\usepackage{amsfonts}
\usepackage{txfonts}
\usepackage{amssymb}
\usepackage{amsbsy} %for sigma in bold face
\usepackage{epsfig}

\def\be{\begin{equation}} \def\ee{\end{equation}}
\def\bea{\begin{eqnarray}} \def\eea{\end{eqnarray}}

\def\nn{\nonumber}

\def\bsigma{{\boldsymbol \sigma}}

\def\bQ{{\bf Q}}
\def\bk{{\bf k}}

\def\bx{{\bf x}}

\def\bp{{\bf p}}

\def\bK{{\bf K}}

\def\la{\langle}
\def\ra{\rangle}

\def\rw{\rightarrow}

\begin{document}

\title{Fractionalized Metals and Superconductors in Three Dimensions}

\author{Zhong Wang}
%\altaffiliation{wangzhongemail@gmail.com}
\affiliation{ Institute for
Advanced Study, Tsinghua University, Beijing,  China, 100084}

\affiliation{Collaborative Innovation Center of Quantum Matter, Beijing 100871, China }

\date{ \today}

\begin{abstract}

We study three-dimensional metals with nontrivial correlation
functions and fractionalized excitations. We formulate for such
states a gauge theory, which also naturally describes the fractional
quantization of chiral anomaly.  We also study fractional
superconductors in this description. This formulation leads to the
``three-dimensional chiral Luttinger liquids'' and fractionalized
Weyl semimetals, which can arise in both fermion and boson models. We
also propose experiments to detect these fractionalized phases.

\end{abstract}

\pacs{73.43.-f, 71.10.Hf, 74.20.Mn, 71.45.-d, 71.70.-d}

\maketitle

\emph{Introduction.}

Fractionalization of quantum numbers
\cite{jackiw1976,su1979,su1980,heeger1988,fradkin1983,laughlin1999,kivelson2001,
fradkin2013,luther1974,steinberg2008,le2008,qi2008b,ran2008,qi2008A,
lee2007,chamon2008,senthil2004,seradjeh2009,bramwell2001,castelnovo2011}
such as electrical charge is an intriguing phenomenon in condensed
matter physics. Fractionalization in weakly interacting systems is
usually associated with topological solitons, a simplest example
being the Su-Schrieffer-Heeger
soliton\cite{su1979,su1980,heeger1988,fradkin1983} in polyacetylene.
Fractionalized excitations can also arise in strongly correlated
systems, in which strong interaction among electrons plays an
essential role in fractionalization. In fact, one of the best known
examples of charge fractionalization is the elementary excitations
(quasi-particles or quasi-holes) carrying charge $\pm e/m$ in the
$m$-th Laughlin state\cite{laughlin1983,tsui1982,klitzing1980}, which
describes a fractional quantum Hall state,  one of the prototypes of
strongly correlated systems. The one-dimensional (1D) chiral
Luttinger liquids\cite{wen1990,wen1990a,wen1992,chang1996}  (CLLs) at
the edge of the Laughlin states are fractionalized metals, for which
the electron Green's function follows a nontrivial
power-law\cite{wen1990,wen1990a,wen1992,chang1996} \bea G
(x,t)\propto \frac{1}{(x-v_F t)^m} \eea where $v_F$ is a velocity parameter. This correlation is in sharp contrast
with expectation of Landau's Fermi liquid theory.

Charge fractionalization is a much more subtle problem for strongly
correlated systems in spatial dimensions higher than one ($d>1$), due
to absence of powerful analytical tools such as bosonization.
Describing fractionalized metals in $d>1$ analogous to the 1D chiral
Luttinger liquids is one of the motivations of the present work.
A deeper motivation, as we now explain, is to describe a class of quantum phenomena we dubbed ``fractional anomalies''. This was the initial motivation of this work.

In quantum mechanics, an ``anomaly'' refers to the failure of a
symmetry of the classical action to be a symmetry of the full quantum
theory. The earliest example is the chiral
anomaly\cite{peskin1995,adler1969,bell1969}, which implies chiral
current nonconservation in the presence of nontrivial gauge field
backgrounds. Remarkably, chiral anomaly also has significant implications in condensed matter systems\cite{qi2008,volovik2003,wang2011b,ryu2012, wang2014a,wang2013b,son2012},
where it is deeply related to topological states such as topological
insulators and topological
superconductors\cite{qi2010a,hasan2010,qi2011,kane2005a,
bernevig2006a,kane2005b,bernevig2006c,
konig2007,fu2006,moore2007,qi2008,fu2007b,roy2009a,
wang2012a,schnyder2008,kitaev2009,wang2010a,
chen2012,wang2010b,vishwanath2013,xu2013a}. As a simplest example,
the one-dimensional edge of 2D integer quantum Hall edge states has
the chiral anomaly for electrical current\footnote{We have the factor
$e^2$ here because we are concerned with electrical current. For the
current associated with particle number, the factor is $e$ instead of
$e^2$.}  \bea \partial_\mu j^{\mu R}= n_R\frac{
e^2}{4\pi}\epsilon^{\mu\nu}F_{\mu\nu} \eea  the superscript and
subscript `$R$' appears because low energy fermions have definite
chirality, which is taken to be right-handed, i.e., they all move
towards the right direction. Here the integer $n_R$ is the number of
these edge modes. According to the bulk-edge correspondence, $n_R$ is
also equal to the Chern number of the two-dimensional bulk of quantum
Hall insulators. In this example the origin of chiral anomaly is
transparent: The transverse current (Hall current) in the bulk adds
or removes charges at the edge, which is regarded as charge
noncnservation by an edge observer.

On one hand, there have been various arguments pointing to the
quantization (non-renormalization) of chiral anomaly. On the ohter
hand, the existence of fractional quantum Hall effects has deepened
our understanding of chiral anomaly. For the Laughlin state with
filling factor $\nu=\frac{1}{m}$, the chiral edge state has the
fractional chiral anomaly\cite{wen2004} $\partial_\mu j^{\mu R} = \frac{1}{m}\frac{
e^2}{4\pi}\epsilon^{\mu\nu}F_{\mu\nu}$, which is consistent with the
chiral Luttinger liquid
theory\cite{wen1990,wen1990a,wen1992,chang1996}.

In 3D, the integer chiral anomaly reads $\partial_\mu j^{\mu R}
=n_R\frac{e^3}{32\pi^2}\epsilon^{\mu\nu\rho\sigma}F_{\mu\nu}F_{\rho\sigma}
$ and $\partial_\mu j^{\mu L} =
-n_L\frac{e^3}{32\pi^2}\epsilon^{\mu\nu\rho\sigma}F_{\mu\nu}F_{\rho\sigma}
$, or more compactly\cite{peskin1995}, \bea \partial_\mu j^{\mu 5} =
(n_R+n_L)\frac{e^3}{32\pi^2}\epsilon^{\mu\nu\rho\sigma}F_{\mu\nu}F_{\rho\sigma}
\label{integer-3d} \eea where $n_R$ and $n_L$ is the number of modes of right-handed and left-handed chiral fermions, respectively, and the chiral current $j^{\mu 5}\equiv j^{\mu R}-j^{\mu L}$. Note that in
Eq.(\ref{integer-3d}) we have the factor $e^3$ because $j^{\mu R}$
refers to electrical current (for the current associated with
particle number this factor should be $e^2$).  The chiral anomaly
have interesting physical implications for Weyl semimetals (and Weyl
superconductors) \cite{wan2011,volovik2003,burkov2011,
zyuzin2012a,witczak2012,hosur2012, aji2011,liu2012, xu2011,
wang2012e, halasz2012,
kolomeisky2012,jiang2012,delplace2012,meng2012,
garate2012,grushin2012,son2012,wang2013a,
gorbar2013,liu2014,li2012a, xu2013b} (``Weyl fermions'' is another
name of chiral fermions).

By analogy with 1D fractional chiral anomaly, the ``fractional chiral anomaly'' for right-handed chiral fermions in 3D reads
\bea \partial_\mu j^{\mu R} =\nu\frac{e^3}{32\pi^2}\epsilon^{\mu\nu\rho\sigma}F_{\mu\nu}F_{\rho\sigma} \label{fractional-anomaly} \eea $\nu$ being a non-integer rational number. We shall not repeat the same equation for left-hand chiral fermions. However, such a fractional anomaly can never be achieved in any noninteracting picture or Landau's Fermi liquid theory.  In fact, free fermions always have integer anomaly, and modest interaction cannot renormalize this integer quantization. Invalidating the Landau's Fermi-liquid theory, certain strong interactions may allow fractional quantization, however, a simple formulation of how this actually happens is lacking.
As we shall show, the fractional chiral anomaly is naturally realized in the fractionalized metals studied in this paper.

With these motivations, we studied fractionalized (semi-)metals in 3D
with nontrivial exponents in correlation functions. The elementary excitations carry a fraction of electron charge. In our
formulation an important role is played by the Higgs mechanism, which
gives mass to gauge bosons, such that their effects are suppressed at
low energy. The results of this paper is applicable to 3D ``chiral
Luttinger liquids'' at the surface of 4D quantum Hall states, and to
fractionalized Weyl semimetals, which in principle can be realized in
experiments.

\emph{$SU(N)$ gauge theory formulation of three-dimensional fractionalized metals.}

The Hamiltonian of electron in a periodic lattice system can be
generally written as \bea H=\sum_{ij}  c^\dag_i [t_{ij}e^{ieA_{ij}}+
eA_0(i)\delta_{ij}]c_j + \sum_{ij}  V_{ij}n_i n_j + \dots   \eea
where the subscript $i,j$ refer to all microscopic degrees of freedom
including site, spin, and orbital, $n_i$ is the particle number operator, and $A_0(i)$ and $A_{ij}$ are
temporal and spatial components of a classical electromagnetic
potential added for later convenience.

To investigate possible fractional phases, the
trick\cite{wen1991a,maciejko2010a,swingle2011} we will use is to write the
electron operator $c_i$ in terms of parton operators $q_{ia}$ \bea
c_i=q_{ i1}q_{ i2}\dots q_{ iN} = \frac{1}{N!} \epsilon_{ab\dots
c}q_{ia}q_{ib}\dots q_{ic} \eea where the parton operators $q_{ ia}$
($a=1,2\dots N$) are fermionic. The electron operator $c_i$ is
invariant under the local gauge transformation $q_{ a i}\rw
W_{i,ab}q_{ bi}$, where $W_i\in SU(N)$. We will see shortly that this
implies that each parton is coupled to an emergent $SU(N)$ gauge
field\cite{yang1954,wilson1974} with lattice gauge potential
$(a_{0}(i), a_{ij})$ to be defined below. We also note that this parton approach can work if the original fermions $c_i$ are replaced by bosons $b_i$, the only modification being that $N$ becomes even integer.

Within the mean field approximation, the quadratic Hamiltonian for
the partons reads \bea H_{{\rm mean}} && =\sum_{ij} \sum_{ab}  (
t_{ij}K_{ij,ab} e^{ieA_{ij}/N}q_{ia}^\dag q_{jb} + V_{ij}K'_{ij,ab} e^{ieA_{ij}/N}q_{ia}^\dag q_{jb} \nn \\ &&   +\dots)
+\sum_i\sum_{ab} q_{ia}^\dag [\lambda_{ 0}^l(i) T^l_{ab} +
e\frac{A_0(i)}{N}\delta_{ab}]q_{ib} \label{mean-1} \eea where \bea
K_{ij,aa'} && = e^{ieA_{ij}(N-1)/N}(1/N!)^2 \nn \\ && \la
\epsilon_{ab\dots c}(q_{ib}\dots q_{ic})^\dag \epsilon_{a'b'\dots
c'}(q_{jb'}\dots q_{jc'})\ra, \nn \eea and similarly for $K'_{ij,ab}$. In Eq.(\ref{mean-1}) we have added the Lagrange multiplier $\lambda_{0}^l(i)$
($l=1,\dots, N^2-1$) to ensure the
constraints $q^\dag_{i1}q_{i1}=\dots=q^\dag_{iN}q_{iN}$ for the
physical Hilbert space, $T^l$ being the generators of $SU(N)$.
Writing the mean-field Hamiltonian in Eq.(\ref{mean-1}) more
compactly, we have \bea H_{{\rm mean}} =\sum_{ij} \sum_{ab}
q_{ia}^\dag [U_{ij,ab} e^{ieA_{ij}/N}+ (\lambda_{ 0}^l(i) T^l_{ab} +
e\frac{A_0}{N}\delta_{ab})\delta_{ij}]q_{ jb} \label{mean} \eea where
$U_{ij,ab}= t_{ij}K_{ij,ab} +V_{ij}K'_{ij,ab}+\dots$. It is readily understood
that $ U_{ij,ab}$s are dynamical variables with their effective
Lagrangian $L_{{\rm eff}}(U)$, whose explicit form does not concern
us here. We can write the $N\times N$ matrix $U_{ij}$ as \bea
U_{ij}=\bar{U}_{ij}e^{i  a_{ij}} \eea in which $a_{ij}=a^l_{ij}T^l$ is a $N\times N$
Hermitian matrices describing the
$SU(N)$ `phase' fluctuations of $U_{ij}$ around the mean field value
$\bar{U}_{ij}$, while the amplitude fluctuation of $U_{ij}$ is
ignored. Similarly, we can split the Lagrange multiplier as
$\lambda_0(i)=\bar{a}_{ 0}(i)+a_{0}(i)$, $\bar{a}_{ 0}(i)$  and
$a_0(i)$ being the vacuum expectation value and the fluctuation of
$\lambda_0(i)$, respectively. We can see that $a_0(i)$ can be
regarded as the temporal component of an emergent $SU(N)$ gauge
field, whose spatial components are $a_{ij}$.

Now let us discuss the dynamics of the $SU(N)$ gauge potentials
$(a^l_{0}(i), a_{ij}^l)$ in the long wavelength limit, which are
referred to as ``$a^l_\mu$'' ($\mu=0,1,2,3$), or more compactly as ``$a^l$'' when
there is no confusion. The most important problem now is whether the
gauge bosons are massless or massive, namely, whether the effective
action $L_{{\rm eff}}(U)$ contains mass terms for $a^l$. In fact, the
gauge bosons can become massive by the Higgs mechanism, the vacuum
expectation $(\bar{a}_{ 0}(i),\bar{U}_{ij})$ playing the role of
``Higgs fields''. To be precise, the gauge bosons $a^l$ will
generally be massive if there exists certain loop $C_i$ given by
$i\rw j\rw k\dots\rw l\rw i$ for which $[T^l, P(C_i)]\neq 0$, where the flux
$P(C_i)\equiv  \bar{U}_{ij} \bar{U}_{jk}\dots \bar{U}_{li}$,
$\hat{P}$ being the path ordering. According to this general
criterion, there are several scenarios\cite{wen2004} associated with
different types of $(\bar{a}_{ 0}(i),\bar{U}_{ij})$:

\begin{enumerate}

\item
``Generic flux''. For any $SU(N)$ group generator
$T^l$, there exists certain loop $C_i$ for which the commutator
$[P(C_i), T^l]\neq 0$. All gauge bosons are massive in this case.

\item
Trivial flux. We have $P(C_i) =1$ for an arbitrary loop $C_i$, in other words, $[P(C_i), T^l]=0$ for all $T^l$ and all $C_i$. All gauge bosons are massless in this case.

\item
Coplanar flux. For some of of the $SU(N)$ group generators, say $T^m$, the relation
$[P(C_i), T^m]= 0$ ( for any $C_i$ ) is satisfied; while for other
group generators, say $T^n$, there exists at least one choice of
$C_i$ such that $[P(C_i), T^n]\neq 0$. The gauge bosons associated
with the first class of group generators are massless, while those
associated with the second class are massive.

\end{enumerate}
Similar scenarios emerge in the $SU(2)$ gauge-field formulation of
spin liquid\cite{wen2004,wen1991b,lee2006}. Without going into technical details, here
we simply provide an intuitive argument for these scenarios (not a
proof). In the continuum limit, the effective Lagrangian generally
contain a term ${\rm tr}([a_\mu, \bar{a}_\nu]^2)$, where $\bar{a}_\nu$ denotes the background field determined by $(\bar{a}_{ 0}(i),\bar{U}_{ij})$.  When this term is
non-vanishing, it can be regarded as a mass term for $a_\mu$, therefore, the Higgs
mechanism of non-Abelian gauge field does not require matter fields
(partons in our context); gauge fields themselves can trigger the
Higgs mechanism because they carry gauge charge\cite{wen2004} (Unlike
the Abelian gauge field theory, in which the gauge boson is charge
neutral).

Here let us focus on the first scenario, namely that all $SU(N)$
gauge bosons acquire an mass from the Higgs mechanism. An explicit ansatz realizing this scenario will be given shortly. In this case
the gauge bosons can only mediate short-range interactions, thus they
do not cause infrared divergences.

If the original electron system has $n$ bands, namely, there are $n$
microscopic electronic states (referred to as $\alpha,\beta,\dots$)
in each unit cell, then the mean-field Hamiltonian Eq.(\ref{mean})
for partons has $nN$ bands.

Suppose that the parton mean-field Hamiltonian $H_{{\rm mean}}$ has
$m$ valleys near $\bK_s$ ($s=1,2,\dots,m$), where the band structure
is that of right-handed chiral fermions (Weyl fermions). Within the
valley around $\bK_s$, the two low energy bands are described
approximately by the Hamiltonian $q^\dag(\bk) \hat{P}_s
h_s(\bk)\hat{P}_s q(\bk)$, where $q(\bk)$ is the abbreviation for
$(q_1(\bk), q_2(\bk),\dots, q_{nN}(\bk))^T$, $\hat{P}_s$ is the
projection operator to the two low energy bands, and  \bea h_s(\bk)
=\sum_{i,j=1,2,3} v_{ij} \sigma_i (\bk-\bK_s)_j -\mu \label{parton-H}
\eea where $v_{ij}$ ($i,j=1,2,3$ or $x,y,z$) play the role of Fermi
velocities, and the Pauli matrices $\sigma_i$ ($i=1,2,3$) act on the
two low energy bands. In the following we will take all $\bK_s=0$ and
$v_{ij}=v_F\delta_{ij}$ ($v_F>0$) without affecting main physical
conclusions.  At $\mu=0$, the Fermi surface shrinks to a point. By
dimensional analysis, various short range interactions such as
$q^\dag  q^\dag  q q$ (indices omitted) are irrelevant in the
renormalization group sense. The electromagnetic interaction is
marginally irrelevant\footnote{It can generate logarithmical
corrections to the power-laws of fermion Green's functions, which we
shall ignore at this stage.}. Therefore, at low energy or long
distance the Green's function of partons is just that of the free
fermions. The elementary excitations are partons (and holes of
partons) carrying charge $\pm e/N$, which is a direct manifestation
of fractionalization.

In the momentum space, the parton Green's function is denoted as
$g_{\alpha a,\beta b}(\omega,k)$, where $\alpha,\beta=1,2,\dots,n$
and $a,b=1,2,\dots,N$. In the low energy limit it is given by $g=
\sum_s \hat{P}_s g_s(\omega,\bk) \hat{P}_s$, with $g_s(\omega,\bk)$ being
the contribution of the $s$-th valley. In our simplest ansatz
it reads $g_s(\omega, k)=1/(\omega+i 0^+{\rm sgn}(\omega) -
v_F\bsigma\cdot\bk)$. Fourier-transformed to the real space, the
parton Green's function is \bea g_s (\bx,t) \propto \frac{v_F t +
\bsigma\cdot{\bf x}}{(x^2-v_F^2 t^2)^2} \label{real-space} \eea where $x^2\equiv
x_1^2+x_2^2+x_3^2$. The electron Green's function is \bea
G_{\alpha\beta}(\bx,t) = \det g_{\alpha a, \beta b}(\bx,t)  \label{det}
\eea where the determinant is calculated regarding the color index $a$ ($b$ ) as the
row (column) index of matrix $g_{\alpha a,\beta b}$,  the
indices $\alpha,\beta$ being fixed. Because $g(\bx,t) \propto 1/t^3$ in
the long time limit, it readily follows that in this limit \bea G(0,t)\propto
\frac{1}{t^{3N}} \label{power} \eea Similarly, we have
$G(\bx,0)\propto 1/x^{3N}$ in the long distance limit. The power-law
behavior of Green's function, with the exponent quantized as an
integer\cite{note-1}, is reminiscent of the chiral Luttinger liquids in 1D. As a
comparison, the 1D chiral Luttinger liquid at the edge of $m$-th
Laughlin state has\cite{wen2004} $G(0,t)\propto  t^{-m}$.

Since each parton carries electrical charge $e/N$, the coefficient of
chiral anomaly in Eq.(\ref{fractional-anomaly}) is \bea \nu=
(1/N)^3\times m = m/N^3 \label{nu} \eea which can be readily obtained
from the triangular diagram\cite{peskin1995}, with parton propagators
extracted from Eq.(\ref{parton-H}). Due to the non-renormalization
property of chiral anomaly, modest modifications of Hamiltonian
cannot change this quantized $\nu$.

%
%Eq.(\ref{det}), Eq.(\ref{power})
%and Eq.(\ref{nu}) are among the central equations in this paper.

Sofar we have formulated a self-consistent theory of fractionalized
metals in 3D, for which the Green's function follows a nontrivial
power-law, and the chiral anomaly is fractionally quantized. Now we
would like to discuss the physical applications of this formulation.
The most direct application can be found at the surface of 4D
fractional quantum Hall
effects\cite{zhang2001,bernevig2002,li2012,wang2014}.  From our
formulation it is clear that fractionalized metal (``3D chiral
Luttinger liquid'') is a possible and consistent scenario in certain
regime. Our formulation can also be applied to Weyl semimetals with
equal number of modes of right-handed and left-handed chiral fermions. For Weyl
semimetals the nontrivial exponents in Green's function can be
measured in tunneling experiments, which provide information of the electron density of states.

Now let us study two simple examples. We consider single-band lattice boson models, and write the (hard-core) boson operator $b_i$ as $b_i=q_{i1}q_{i2}$. The mean-field ansatz reads $H_{{\rm mean}}= \sum_{\bk} q^\dag h(\bk) q$, $q$ being the shorthand notation for $(q_1(\bk),q_2(\bk))^T$.  The first example we consider is $h(\bk)=v_F  \bsigma\cdot\bk$, which gives nonzero masses to all three $SU(2)$ gauge bosons because none of $\sigma_i$ ($i=1,2,3$) commutes with $v_F\bsigma\cdot\bk$ for a generic $\bk$. According to Eq.(\ref{nu}), we have fractional anomaly \bea \nu=(1/N)^3=1/8 \eea Such a fractional anomaly can be realized at the boundary of a 4D (boson) quantum Hall insulators. The parton propagator for this model is exactly given by Eq.(\ref{real-space}). The boson Green's function can be obtained from Eq.(\ref{det}), with the simplification in this special case that the indices $\alpha,\beta$ are absent. Explicitly, we have   \bea G(\bx,t)=\det g \propto \frac{1}{(x^2-v_F^2t^2)^3} \eea which implies $G(0,t)\propto 1/t^6$ in the $t\rw\infty$ limit.

The second example of parton mean-field Hamiltonian is $ h(\bk)=[2t_x(\cos k_x-\cos K)+m(2-\cos k_y -\cos k_z)]\sigma_x +2t_y\sin k_y\sigma_y +2t_z\sin k_z \sigma_z $, whose form is borrowed from Ref.\cite{yang2011}. It has two Weyl valleys near $\pm\bK=(\pm K, 0, 0)$, where $h(\bp)\approx \mp 2t_x (\sin K ) p_x\sigma_x +2t_y p_y\sigma_y +2t_z\sigma_z$, with $\bp\equiv \bk\mp \bK$. The elementary excitations are chiral fermions carrying charge $\pm e/2$, and the fractional anomaly is given by Eq.(\ref{fractional-anomaly}) with $\nu=1/8$. There is a similar equation for $j^{\mu L}$ with $\nu=-1/8$.  The boson Green's function reads $G(\bx,t)=\det g$, which scales as $G(0,t)\propto 1/t^6$. This example can be realized in 3D lattice boson model (rather than just realized at the surface of 4D lattice models).

\emph{One dimensional fractionalized chiral modes propagating along dislocations.}

\begin{figure}
\includegraphics[width=5.0cm, height=4.0cm]{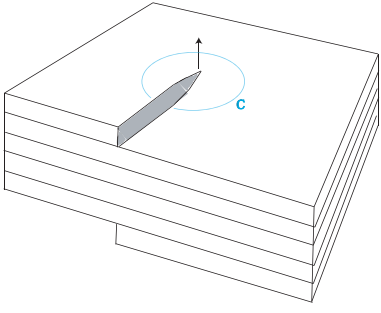}
\caption{Dislocation of density wave. A screw dislocation $l$ is
parallel with the arrow. Along $l$ the one-dimensional fractionalized
modes propagate unidirectionally. The ``Hall conductance'' of this chiral mode is $e^2/N^2h$, which signifies fractionalization (see main text). } \label{dislocation}
\end{figure}

One way to detect the charge fractionalization in the fractionalized Weyl semimetals is to induce an energy gap and then create dislocations.

For simplicity, let us suppose that the mean field Hamiltonian
Eq.(\ref{mean}) has one valley (right-handed fermion) around $\bK_R$,
where the low energy Hamiltonian is $h_R(\bk)=
v_F\bsigma\cdot(\bk-\bK_R)$, and another valley (left-handed fermion)
at momentum $\bK_L$, described by
$h_L(\bk)=-v_F\bsigma\cdot(\bk-\bK_L)$. Now let us add an external
magnetic field ${\bf B}=B\hat{z}$. In the presence of this magnetic
field, the energy spectra are $E_n(p_z)=\pm v_F \sqrt{p_z^2+2eB
l/N}$ with $l=0,¡¡1,¡¡2,\dots$, where $p_z=(\bk-\bK_{R/L})_z$.
For the $l=0$ mode, the $\pm$ sign before $v_F$ is determined by the
chirality ($+$ for right-handed, $-$ for left-handed). Therefore, the¡¡effective Hamiltonian for the zeroth Landau level can be written
compactly as  $h_{0}(p_z)= v_F\tau_z p_z$, where the Pauli matrix
$\tau_z$ refers to the two chiralities. Note that $h_{0}(p_z)$ is
independent on $B$. Due to perfect nesting of Fermi surface, an
infinitesimal interaction can dynamically generate a
mass\cite{yakovenko1993,yoshioka1981,wang2013a,yang2011} $m(r) = \la
q^\dag ({\bf x})\tau_+ q ({\bf x})\ra \propto e^{ i [\bQ\cdot{\bf
x}+\theta({\bf x})]}$, where $\tau_+=\tau_x+i\tau_y$,
$\bQ=\bK_L-\bK_R$, and $\theta({\bf x})$ is a slowly varying
variable. The (charge or spin) density vary as\cite{wang2013a}
$\cos(\bQ\cdot{\bf x}+\theta({\bf x}))$, in which $\theta({\bf x})$
determines the locations of peaks and troughs of the density wave.

Suppose that there is a line dislocation $l$, such that for a loop
$C$ around $l$ we have $\int_C d\theta =2\pi$, thus the peaks and
troughs shift by one wavelength by making a circle around $l$  [see
Fig.\ref{dislocation}]. In Ref.\cite{wang2013a}, it was shown for a
similar problem (integer case) that there is a chiral mode along $l$, which is
analogous to the edge state of integer quantum Hall effect with Hall
conductance $e^2/h$. By similar calculation, it can be shown for the
present problem that there is chiral modes along $l$ with ``Hall
conductance'' $e^2/N^2 h$. In this sense the fractionalized chiral
mode along $l$ is analogous to the 1D chiral Luttinger liquid. Measuring this fractional ``Hall conductance'' will be one way to confirm the fractionalized metals in 3D.

\emph{Fractional superconductors in 3D.}

In the previous sections we have focused on the case with parton chemical
potential $\mu=0$. When $\mu\neq 0$, an infinitesimal attractive
interaction can induce Cooper instability, and the ground state is a
superconductor. Let us consider a Weyl valley around $\bK_s$. For a
short-range interaction $g\delta^{(3)}({\bf x})$, the superconducting
gap is \bea \Delta_{{\rm s-wave}} = \la
q_{R\uparrow}(\bK_s+\bp)q_{R\downarrow}(\bK_s-\bp)\ra \propto
e^{-c/g\rho} \eea where $\uparrow$ and $\downarrow$ refer to the two low
energy degrees of freedom, $\rho$ is the density of states of partons at the Fermi level, and $c$ is a numerical coefficient of
order of unity. More complicated scenarios such as color
superconductors\cite{rajagopal2000,anglani2014} can also arise in our
description, which will be left for future studies. Here we would
like to focus on model-independent physical consequences.

One of the physical predictions is about the Josephson
effect\cite{josephson1962,anderson1963}. In fact, the parton
Cooper pairs carry charge $2e/N$, therefore, we expect fractional
Josephson effects. In the presence of a voltage $V_0$ between two
fractional superconductors connected by a weak link, alternating
tunneling current with frequency \bea \omega_0 =\frac{2eV_0 }{N\hbar}
\eea can be observed (In this formula we have restored the Planck
constant $\hbar$, which has been set to unity in our previous
presentation). We would like to mention that the fractional
Josephson effect has also been studied in 2D
systems\cite{cheng2012,vaezi2013,lindner2012,clarke2013} in different
approaches.

\emph{Conclusions.}

In the present paper we have studied fractionalized (semi-)metals in
3D through an $SU(N)$ gauge theory. We have found for them power-law
Green's functions with quantized exponents. This formulation will be
useful to 3D chiral Luttinger liquids and fractionalized Weyl
semimetals. This gauge-theoretical formulation resembles the standard
model of particle physics, in which the chiral fermions are coupled
to $U(1)$ and $SU(N)$ gauge fields with $N=2,3$, the $SU(2)$ being
suppressed at low energy due to the Higgs mechanism. In the field of
condensed matter, we believe that \emph{fractional anomalies} will provide
much information about fractional topological states of quantum matter. In addition to the significance of their own right, the fractionalized (semi-)metals can also be regarded as the ``mother states'' of gapped fractional topological states. In
future it will be fruitful to establish explicit many-body Hamiltonians
for the scenarios proposed in the present work, and to explore the implications of fractional anomalies in depth.

The author would like to thank Xiao-Gang Wen, Yi Li, Congjun Wu, and
especially Shou-Cheng Zhang for insightful discussions. The author is
supported by NSFC under Grant No. 11304175 and Tsinghua University
Initiative Scientific Research Program ( No. 20121087986).

\bibliography{fractional_3}

\end{document}